\documentclass{article}

     \usepackage[nonatbib, final]{neurips_2019_ml4ps}
     \usepackage[pdftex]{graphicx}
     
    \newcommand{\arcsec}{^{\prime\prime}}
    \usepackage{xcolor}

\usepackage[utf8]{inputenc} 
\usepackage[T1]{fontenc}    
\usepackage{hyperref}       
\usepackage{url}            
\usepackage{booktabs}       
\usepackage{amsfonts}       
\usepackage{nicefrac}       
\usepackage{microtype}      
\usepackage{graphicx}

\title{Single-Frame Super-Resolution of Solar Magnetograms: 
Investigating Physics-Based Metrics \& Losses}

\author{%
  Anna Jungbluth \\
  University of Oxford \\
  \texttt{anna.jungbluth@physics.ox.ac.uk}\\
  \And
  Xavier Gitiaux \\
  George Mason University \\
  \texttt{xgitiaux@gmu.edu} \\
  \And
  Shane A.~Maloney \\
  Trinity College Dublin / \\
  Dublin Institute for Advanced Studies \\
  \texttt{shane.maloney@tcd.ie} \\
  \And
  Carl Shneider \\
  Center for Mathematics \\
  and Computer Science (CWI) \\
  \texttt{carl.shneider@cwi.nl} \\
  \And
  Paul J.~Wright \\
  Stanford University \\
  \texttt{pjwright@stanford.edu} \\
  \And
  Alfredo Kalaitzis \\
  Element AI \\
  \texttt{freddie@element.ai} \\
  \And
  Michel Deudon \\
  Element AI \\
  \texttt{michel.deudon@elementai.com} \\
  \And
  Atılım Güneş Baydin \\
  University of Oxford \\
  \texttt{gunes@robots.ox.ac.uk} \\  
  \And
  Yarin Gal \\
  University of Oxford \\
  \texttt{yarin@cs.ox.ac.uk} \\
  \And
  Andr\'es Mu\~noz-Jaramillo \\
  Southwest Research Institute \\
  \texttt{amunozj@boulder.swri.edu} \\
}

\begin{document}

\maketitle

\begin{abstract}

Breakthroughs in our understanding of physical phenomena have traditionally followed improvements in instrumentation. 
Studies of the magnetic field of the Sun, and its influence on the solar dynamo and space weather events, have benefited from improvements in resolution and measurement frequency of new instruments.
However, in order to fully understand the solar cycle, high-quality data across time-scales longer than the typical lifespan of a solar instrument are required.
At the moment, discrepancies between measurement surveys prevent the combined use of all available data.
In this work, we show that machine learning can help bridge the gap between measurement surveys by learning to \textbf{super-resolve} low-resolution magnetic field images and \textbf{translate} between characteristics of contemporary instruments in orbit.
We also introduce the notion of physics-based metrics and losses for super-resolution to preserve underlying physics and constrain the solution space of possible super-resolution outputs.

\end{abstract}

\section{Introduction}

For 50 years, a series of instruments have provided images of the Sun's magnetic field (known as \emph{magnetograms}) to study the field's origin and evolution. However, differences in resolution, noise properties, saturation levels, spectral inversion techniques, and other instrument specifics introduce inhomogeneities that prevent direct comparison of magnetograms across instruments.  The calibration and homogenization of magnetograms remains an unsolved problem today. 

Traditionally, magnetogram calibration has been performed through pixel-to-pixel comparison \cite{2012SoPh..279..295L}, histogram equalization \cite{2014SoPh..289..769R}, or harmonic scaling \cite{2019A&A...626A..67V}. Recent advances in deep learning highlight the potential to improve and simplify these calibration processes to build a consistent, high-resolution dataset of magnetograms. Deep neural networks can transform and super-resolve natural images \cite{8723565}, and have recently been successfully applied to magnetograms \cite{mcgregor2017,2018A&A...620A..73A,2018A&A...614A...5D,2019NatAs...3..397K}.

In this work, we use deep neural networks to super-resolve and cross-calibrate magnetograms taken by the Michelson Doppler Imager (MDI) \cite{1995SoPh..162..129S,1995SoPh..162....1D} and the Helioseimic and Magnetic Imager (HMI) \cite{2012SoPh..275..207S,2012SoPh..275..229S,2012SoPh..275....3P}. During its tenure from 1995 -- 2011, MDI acquired magnetograms every 96 minutes, with a detector of $1024 \times 1024$ pixels imaging the Sun at $2\arcsec$ (arcseconds) per pixel. In comparison, HMI, which has been operational since 2010, images the Sun every 45 seconds at $0.5\arcsec$ per pixel with a detector of $4096 \times 4096$ pixels. 

MDI and HMI observations overlap from 2010 to 2011, resulting in nearly $9000$ co-temporal observations of the same physical structures. By training a deep neural network on a subset of observations, MDI magnetograms are super-resolved and upsampled by a factor of four to the size of HMI observations. 
We show that penalty terms based on image gradients in the loss function better preserve the physical properties of magnetograms in the super-resolved reconstructions.

\section{Data processing}

\begin{figure}[!t]
  \centering
  \includegraphics[width=0.9\linewidth]
  {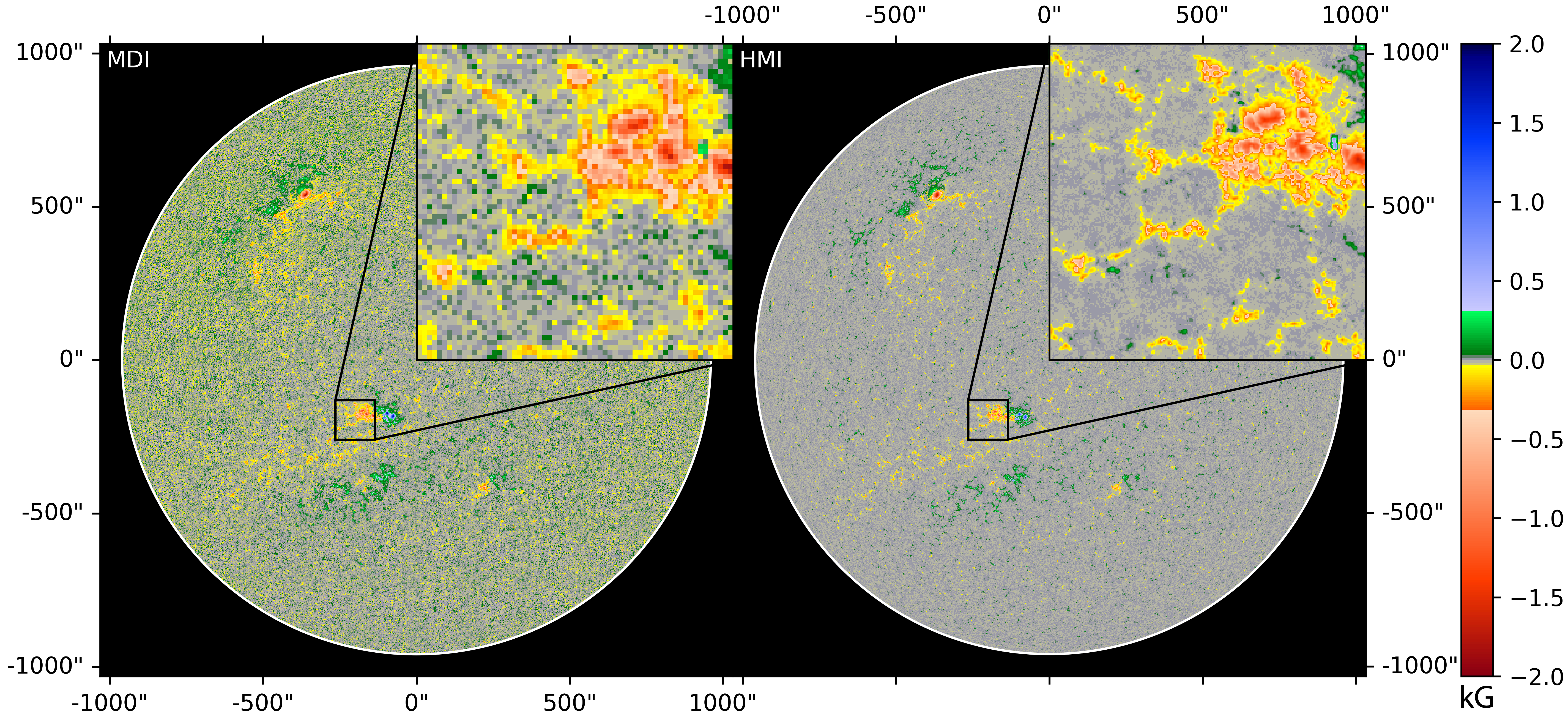}
  \caption{
    Comparison of co-temporal, co-aligned magnetograms obtained by MDI ({\it left}), and HMI ({\it right}). Both images show the full solar disk, re-scaled as if they were observed from $1$~Astronomical Unit, and plotted over the range of $\pm2000$ Gauss.
    The $128\arcsec \times 128\arcsec$ insets were registered using vertical and horizontal shifts to account for the distortion between instruments.
    \label{fig:1}
    }
\end{figure}

While neural networks can learn systematic differences between instruments, random positional, rotational, and temporal differences between images significantly degrade the model's ability to learn the mapping and have to be removed before training. We therefore apply a series of operations to the MDI and HMI datasets:
(1) rotate solar north to the image north to standardize the Sun’s orientation in each image, using instrument pointing information contained in the magnetogram headers;
(2) scale the image to standardize the detector resolution and the size of the Sun in the detector as if it was observed from 1 Astronomical Unit (AU; the average Sun-Earth distance);
(3) register and shift images taken at different times to more closely align features.
The effect of parallax, which is less significant than the effect of temporal discrepancies between instruments, is also removed during the above processing steps.
Fig~\ref{fig:1} shows rotated, distance-corrected, and temporally aligned full-disk magnetograms. 
The resulting $9000$ image pairs are split by randomly selecting one month for validation and testing, and allocating the remaining ten months to training. A monthly split was chosen over a random allocation of images to preserve some of the temporal evolution of the magnetic field during training. In addition, a random allocation of images can lead to many instances where the model is trained and tested on magnetograms spaced only minutes apart. During that time, neither the instrument's perspective nor the magnetic field features have changed enough to avoid a biased evaluation of the model's performance.
The full disk magnetograms are broken into $1024$ $128\arcsec \times 128\arcsec$ patches that are augmented with N-S, E-W reflections and polarity flips during training.

\section{Super-resolution}

{\bf Single-image super-resolution (SISR)} 
approximates the inverse of the blur and down-sampling operators that exist between a high-resolution (HR) image and its low-resolution (LR) counterpart \cite{8723565}.
Super-resolution is an ill-posed problem, as multiple SR outputs can explain a LR input. Compared to classical upsampling methods, such as linear or bicubic upsampling, that do not add new information, learning-based approaches can constrain the non-trivial solution space by exploiting the spatial regularities within a specific distribution of images---in this work, that of magnetograms. 

{\bf Convolutional Neural Networks (CNN)} are the current state-of-the-art in capturing spatial regularities in complex natural images.
With a sufficient amount of annotated examples,
a CNN can learn salient LR-to-HR correlations that best describe the data \cite{SRCNN, SRGAN}. CNNs have successfully been applied to super-resolution tasks in remote sensing, medical imaging and microscopy
\cite{Martens2019, tao2019super, pham2017brain, rivenson2017deep}.
Typical loss functions for super-resolution tasks---like either the $l_1$ or the mean squared error (MSE, $l_2$)---encode the pixel-level error between the SR reconstruction and the HR ground-truth \cite{farsiu2004fast}.

In this work, we use HighRes-net \cite{anonymous2020highresnet}\footnote{\url{https://github.com/ElementAI/HighRes-net}.},
a state-of-the-art architecture for multi-frame super-resolution (MFSR) that recursively fuses an arbitrary number of LR images of the same view.
While our current work focuses on single-frame input images, future work will expand to MFSR to leverage the full capability of the model. Future work will also focus on comparing the performance of HighRes-net to competitive generative models like \emph{Generative Adversarial Networks} (GANs), and \emph{Variational Auto Encoders} (VAEs).
For now, the model's output is compared to a bicubic upsampling baseline. Since bicubic upsampling is a trivial operation, it serves as an ideal "no-effort" baseline to compare the model's output to.

\section{Loss functions \& performance metrics}

Any application of super-resolution within the physical sciences would be amiss if during training, the model had no incentive to preserve the underlying physics of the task at hand\footnote{For instance, zero divergence of the magnetic field needs to be preserved.}. The need for physical accuracy allows us to further constrain the SR solution space. While typical \emph{reconstruction}-based losses (e.g. MSE) suffice to obtain visually convincing SR reconstructions, quantitative methods are required to evaluate the scientific quality of the model's output. This motivates our investigation of (1) differentiable loss functions for physics-driven training, and (2) post-training performance metrics to assess the quality of the SR outputs.

\subsection{Loss functions}

Compared to natural images, the range of pixel values in magnetograms is not constrained, even though most pixels show close to zero magnetic field. Active regions, with large magnetic fields, occur outside the disk center and are orders of magnitude more rare. This is especially problematic for MDI to HMI conversions, as the year of overlap (2010-2011) occurred at a minimum of the solar cycle with few active regions. 

As a consequence of this uneven representation, the model likely learns to preserve small fields well but struggles to reconstruct active regions. While a simple mean squared error can reconstruct broad features in magnetograms, it fails to reconstruct strong magnetic fields (Fig \ref{fig:2}). This is also due to the MSE sensitivity to large errors, which leads to compensating by over-smoothing the output. This over-smoothing can be offset by using an $l_1$ or \emph{Huber loss} \cite{huber1964} instead of an $l_2$ \emph{loss}. Future work will focus on expanding our investigations to alternative loss functions.

To improve the reconstruction of large pixel values, our current work focuses on investigating loss functions composed of an MSE loss with additional penalty terms.
So far, we have mostly focused on an image gradient-based penalty term, but are investigating adding the {\it Kullback-Leibler divergence} \cite{mackay2003information}. 

\smallskip

{\bf Image gradient-based loss}
With a Sobel kernel \cite{Sobel} we approximate the $x$ and $y$ gradients of the SR output ($I_{SR}$) and the HR ground truth ($I_{HR}$).
By computing the mean squared difference between the pixel gradients (G) of the images, we incentivize the model to preserve both magnetic field values and spatial variations:

 \begin{equation}
     Loss = MSE(I_{SR}, I_{HR}) + n\cdot MSE(G_{SR}, G_{HR})
     \label{eqn:gradient-loss}
 \end{equation}
 
In this case, the weighting $n$ of the gradient term is chosen to be $100$ to allow for equal magnitude contributions of both terms to the overall loss function. 

In computer vision, image gradients are often used for edge detection and texture matching \cite{forsyth2002computer}.
For magnetograms, edge detection can help define sharper edges around non-zero magnetic field regions and texture matching can aid in recovering small-scale details that make up a high-resolution image.

\subsection{Metrics}

\begin{figure}[!t]
    \centering
    \includegraphics[width=1\linewidth]{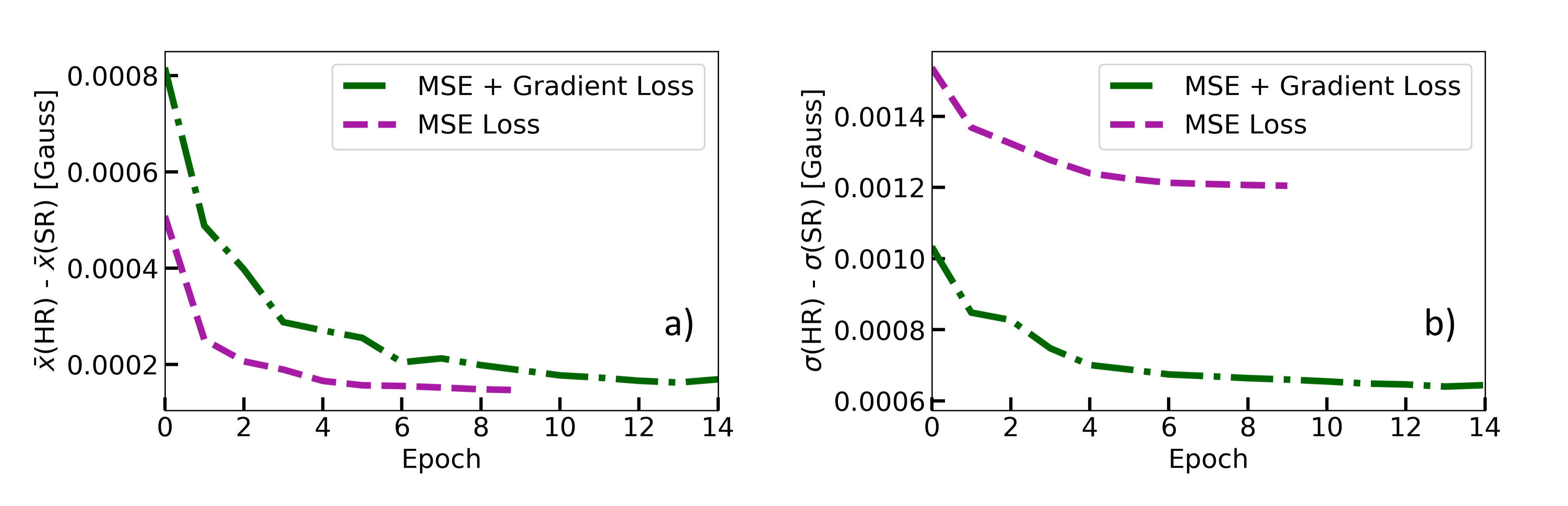}
    \vspace{-2ex}
    \caption{
    Difference between the a) mean and b) standard deviation of the pixel values of the HR HMI target and SR output as a function of training epochs.
    \label{fig:4}
    } 
\end{figure}

\begin{figure}[!b]
    \centering
    \includegraphics[width=0.9\linewidth]{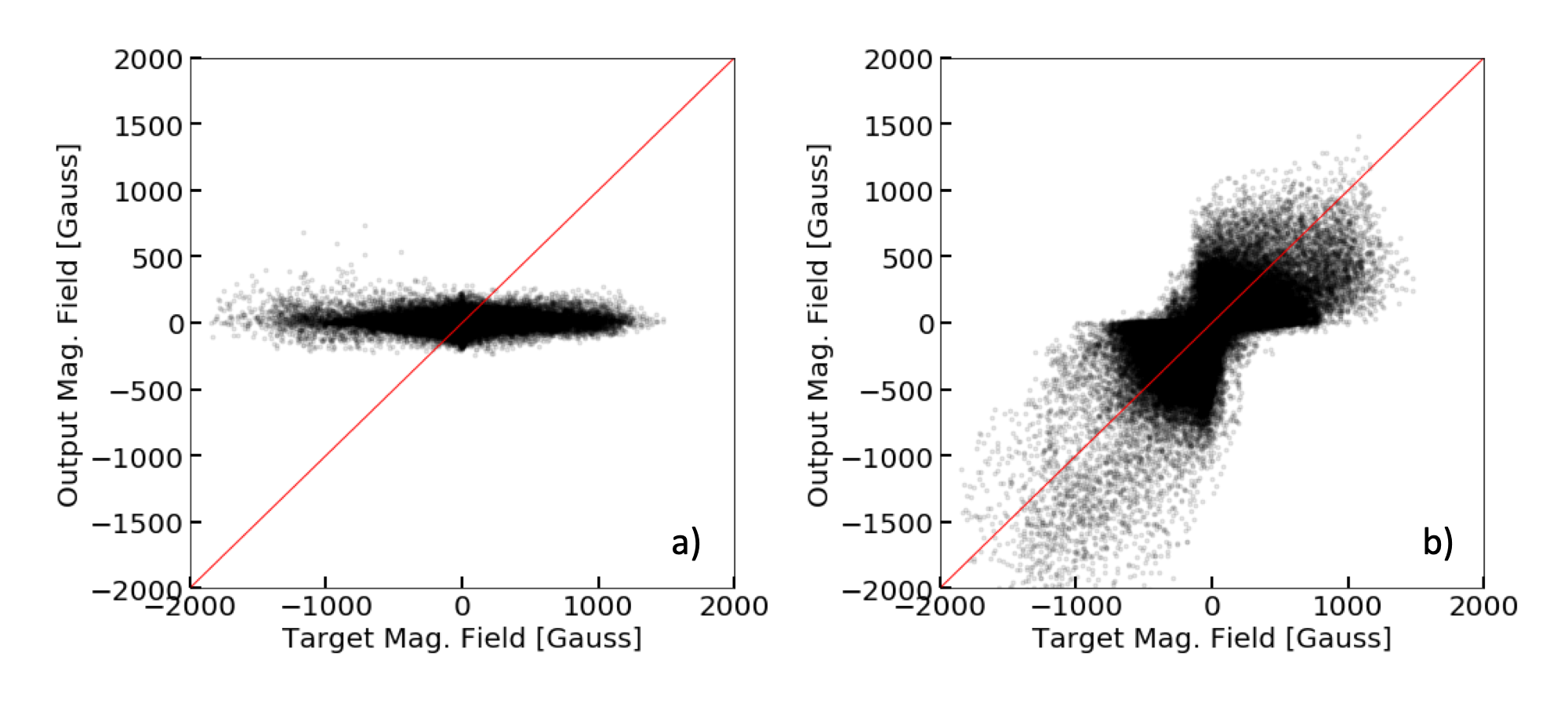}
    \vspace{-2ex}
    \caption{
    Correlation plots of pixel values of the HR HMI target and SR output, using a) an MSE loss and b) an MSE + gradient-based loss function.
    \label{fig:5}
    } 
\end{figure}

Apart from tracking the mean and standard deviation of magnetic field values (Fig \ref{fig:4}), we evaluate the network's performance using three main metrics: (1) correlation plots between the target and output pixel values (Fig \ref{fig:5}); (2) distribution histograms of pixel values to assess the recovery of small and large magnetic fields (Fig \ref{fig:3}-a); and (3) the \emph{information entropy} of an image at multiple scales of spatial binning (Fig \ref{fig:3}-b). The information entropy of a random variable $X$ is defined as $H[X] = -\sum_{x} p(x) \ln p(x)$. By comparing the information entropy of the output and target image, we assess whether the reconstruction contains less detail (over-smoothing) or more detail (hallucination) relative to the ground-truth complexity. At each scale of $2^k$ pixels, for $k \in \{0,1,2,...\}$, we convolve the image with an averaging kernel of size $2^k \times 2^k$ and compute its entropy.

\section{Results}

\begin{figure}[!t]
    \centering
    \includegraphics[width=1\linewidth, trim={0 20ex 0 0}, clip]{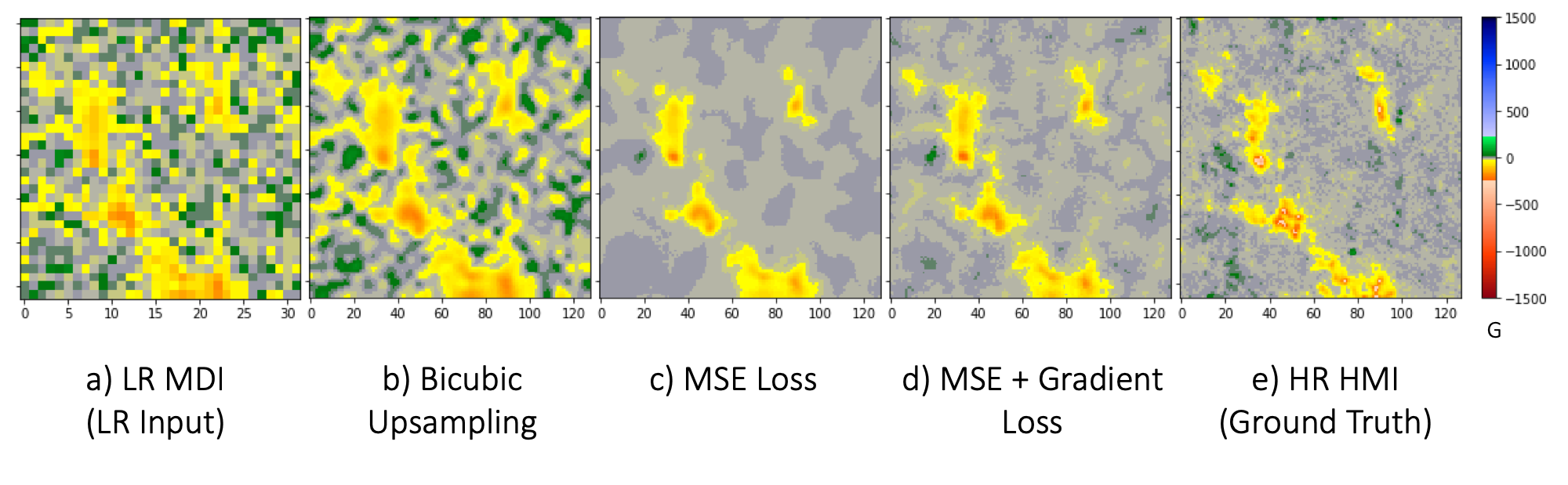}
    \vspace{-1ex}
    \hspace*{7ex}
    {a)} \hfill {b)} \hfill {c)} \hfill {d)} \hfill {e)}
    \hspace*{\fill}
    \caption{
    Comparison of magnetogram patches using different loss functions. Left to right:
    a) LR MDI input,
    b) bicubic upsampling,
    c) MSE loss,
    d) MSE + gradient-based loss,
    e) HR HMI target.
    \label{fig:2}
    }
\end{figure}

\begin{figure}[h]
    \centering
    \includegraphics[width=0.9\linewidth]{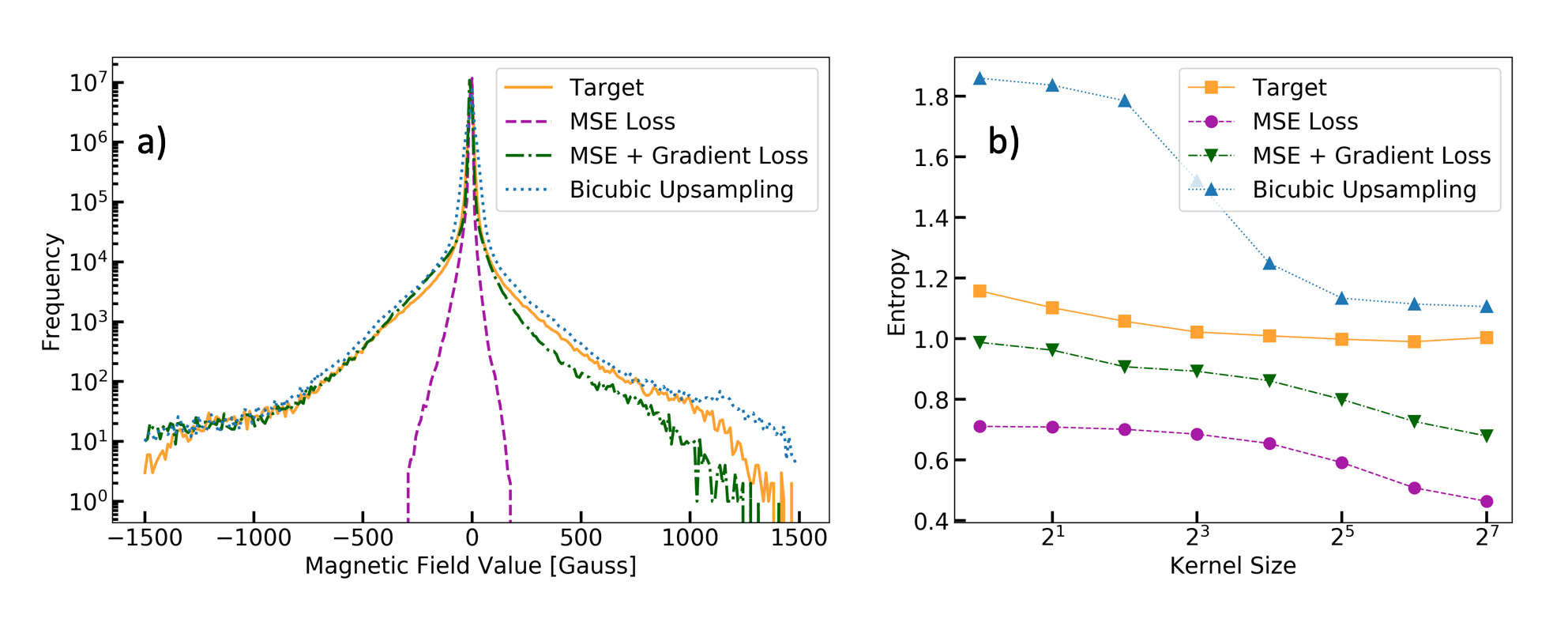}
    \vspace{-2ex}
    \caption{
    a) Pixel histograms and b) multi-scale entropy of the HR HMI target, bicubic upsampling, and the SR output using an MSE, or MSE + gradient-based loss function.
    \label{fig:3}
    } 
\end{figure}

\begin{figure}[!t]
    \centering
    \includegraphics[width=1\linewidth]{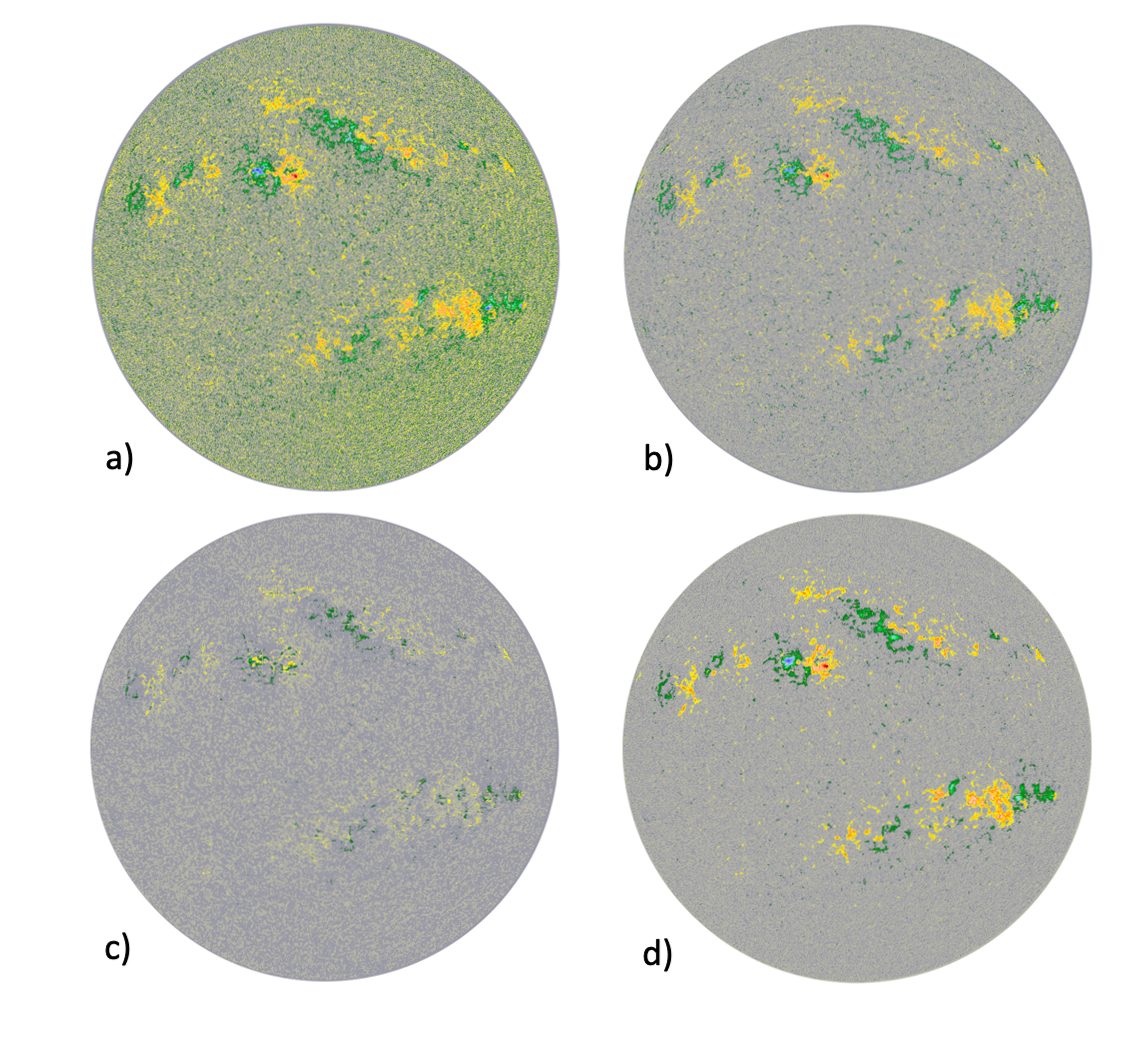}
    \vspace{-2ex}
    \caption{
    Reconstruction of full disk magnetograms; a) LR MDI input, b) HR HMI target, c) SR output using an MSE loss, d) SR output using an MSE + gradient-based loss.
    \label{fig:6}
    } 
\end{figure}

The results of our model, trained with either an MSE or an MSE + image gradient-based loss, are compared to a simple bicubic upsampling of the input magnetograms.
Fig \ref{fig:2} shows the same patch of a magnetogram taken on 2011-04-01 by both MDI or HMI, compared to the bicubic baseline and the model's outputs, and Fig \ref{fig:6} shows the full disk reconstruction. Fig \ref{fig:3}-a shows the distribution histograms of pixel values and Fig \ref{fig:3}-b the multi-scale entropy for all patches of the full disk image.
While bicubic upsampling is able to preserve the pixel distributions well, the quality of the output image is poor (Fig \ref{fig:2}-b). In addition, the multi-scale entropy of the bicubic upsample exceeds that of the target. This may indicate that low-resolution MDI magnetograms contain noise that is not in the high-resolution HMI target. Bicubic usampling extrapolates noise to the high-resolution output. This highlights that bicubic upsampling is not good at instrument conversion because it cannot account for detector specificities.
Using HighRes-net with an MSE loss does well at removing MDI's noise floor (Fig. \ref{fig:2}-c), but fails at reconstructing strong magnetic fields (Fig \ref{fig:3}-a). The correlation plot in Fig \ref{fig:5}-a) confirms the strong clamping of magnetic field values between $-500$ and $500$ Gauss.
In comparison, the composite of the MSE and the image gradient-based loss leads to a substantial improvement in the model's performance. Not only are the pixel distributions of the output and target almost identical, the multi-scale entropies are also more closely matched.
Visually, the features in Fig \ref{fig:2}-d have sharper edges and resemble those of the target image (Fig \ref{fig:2}-e) more closely. Fig \ref{fig:2}-d also shows finer small-scale features, compared to the large, blocky patches of low-magnetic field regions in Fig \ref{fig:2}-c.
Since image gradients are often used for improved edge detection and texture matching, a visual improvement of the magnetograms does not necessarily come as a surprise. Crucially, visuals are not enough to provide evidence of the scientific value of super-resolution for magnetograms. Taking image gradients into account during training improves the model's performance on all presented metrics. Compared to natural images, the image gradients of a magnetogram do not only hold information on texture and feature shapes, but are a direct measure of the magnetic field gradients on the surface of the Sun. Consequently, incentivizing the model to preserve both the magnitude and the gradient of the magnetic field leads to qualitatively and quantitatively promising results for the application of super-resolution to magnetograms.

\section{Conclusions and future work}

We demonstrate the combination of a reconstruction (MSE) and an image gradient-based loss as a way of super-resolving magnetic field images taken by MDI to the resolution of HMI. By adding an image gradient-based penalty term to the loss function, we are able to visually improve the model's output, as well as improve the reconstruction of large magnetic field regions.
These results show the potential of physics-based metrics as regularization terms to constrain ill-defined super-resolution tasks.
Moving forward, we focus on expanding this work to multi-frame super-resolution of all 50 year's worth of magnetogram instruments, paving the way to generate the uniform, high-resolution dataset of magnetograms needed for space weather studies.

\subsubsection*{Acknowledgments}
This work was conducted at the NASA Frontier Development Laboratory (FDL) 2019. NASA FDL is a public-private partnership between NASA, the SETI Institute and private sector partners including Google Cloud, Intel, IBM, Lockheed Martin, NVIDIA, and Element AI. These partners provide the data, expertise, training, and compute resources necessary for rapid experimentation and iteration in data-intensive areas.
P.~J.~Wright acknowledges support from NASA Contract NAS5-02139 (HMI) to Stanford University. 
This research has made use of the following open-source Python packages SunPy \cite{sunpy}, NumPy \cite{numpy}, Pandas \cite{pandas}, and PyTorch \cite{pytorch}. We thank Santiago Miret and Sairam Sundaresan (Intel) for their advice on this project.

\medskip

\small

\bibliographystyle{unsrt}
\bibliography{main_final.bbl}

\end{document}